# Further investigations into the connection between cosmic rays and climate

Dragić A.L. Veselinović N. Maletić D. Joković D. Banjanac R, Udovičić V, Aničin I.
*Institute of Physics, University of Belgrade, Pregrevica 118, Belgrade, Serbia*
dragic@ipb.ac.rs

**Abstract**

Our previous results on the connection between the Forbush decreases (FD) of cosmic-ray intensity and the deviations from the expected values of the diurnal temperature range (DTR) are briefly revisited. The same type of analysis is then extended to the cases of sudden increases of cosmic-ray intensity (GLE), as well as to the search for lattitude effects in the observed correlations. We find that all the investigated correlations appear to manifest both the expected signs and the plausible phase relations, though each one only at the modest confidence level. Moreover, it appears that there is some proportionality between the magnitude of a cosmic-ray intensity change and a corresponding DTR deviation, both in the case of FD and GLE events. Eventual increase of the confidence levels at which these correlations are established would have to wait for the significant increase of the number of well defined and sufficiently intense recorded departures of cosmic-ray intensity from its stationary mean value. On the other hand, the probability of such an accidental multiple coincidence of independent pieces of evidence is certainly very low.

**1. Introduction**

The long-lasting debate about the possible connection between cosmic rays and climate is centered on the plausible assumption that ionization of atmospheric gases by cosmic rays significantly enhances the process of cloud formation, and the controversial evidence in support of the proposition. Some analyses of existing data on global cloudiness and long-term variations of cosmic ray (CR) intensity seem to support the hypothesis[1-3], though the issue is still far from being settled[4-7]. Recent results of the first controlled experiment by the CLOUD collaboration at CERN[12], aimed at revealing the mechanisms behind the aerosol formation in presence of ionizing radiation, encourage further investigation in this direction. Our recent analysis[13] also points at some evidence in favor of the connection. In this previous study we examined the relations between the sudden short-term decreases of cosmic ray intensity, known as the Forbush Decreases (FD), and the universe of deviations from the expected local diurnal temperature range values (the DTR). This work has, however, recently been subjected to some criticism[15,16]. In present work we comment on these criticisms and extend the same type of analysis to the cases of sudden increases of CR intensity, known as the Ground Level Enhancements (GLE). We also check for presence of some latitude dependence of the observed correlations, which is always expected when considering the effects of cosmic rays on Earth.

## 2. Comments on our previous work

Since the present analysis is performed along the same lines as our previous one[13], we shall first try to clarify the methodology used in this earlier work and then comment on the results of the recently repeated analysis along similar lines by Laken et.al.[15] and Erlykin and Wolfendale[16].

The global cloudiness (and the universe of local annual DTR patterns as its proxy) as well as the (low) global intensity of heavily ionizing cosmic-rays, are, in not too extended periods of time, on the average constant. It is difficult to establish in an absolute way the extent to which this on-the-average-constant cosmic-ray intensity contributes to the on-the-average-constant global cloudiness, the CLOUD experiment coming closest to this *experimentum crucis*. Until this is established, we are left with searches for correlations between either long-term periodic or occasional short-term aperiodic departures of these two quantities from their on-the-average-constant values. However, the atmosphere is a noisy system with large fluctuations of the many parameters that describe and influence its state, meaning that the distributions of these parameters are wide, and implying that the occasional non-statistical changes of cosmic-ray intensity have to be higher than some threshold value, sufficient to manifest significant correlation, if any, with the possible corresponding variations of meteorological parameters – the global cloudiness, or the universe of DTR values as its proxy, in our case. The local DTR is defined as the difference between the local maximum and minimum daily temperature, which has been demonstrated to be inversely proportional to local cloudiness[8,9]. To obtain the universe of expected yearly DTR values one has to employ sufficient number of possibly independent local patterns of DTR values for a significant period of years. To accomplish this we performed the superposed epoch analysis on the set of temperature data provided by 210 European meteorological stations for the period of last 50 years. The "universe" in our case thus reduces to Europe, which covers a significant range of latitudes ($35^{o}$ N to $70^{o}$ N) and may hopefully be considered sufficiently representative. On the other hand, as sufficiently representative of the time series of the global cosmic ray flux, we used the neutron monitor data provided by the Mt. Washington observatory, which are available for the period from 1955 to 1990. The 210 average local annual DTR patterns (Fig. 1 of ref.13), define the universe of the expected DTR values for every day of the year at these locations. The distributions of data around average values satisfy normality criteria, what justifies the use of standard Gaussian interval probabilities for estimating the significance of every particular departure from these expected averages. Since the proportion of days when the well isolated non-statistical changes of CR intensity occur, to all the days included into the definition of DTR averages is negligible, these dates are not excluded from the pool of data used for the definition of the DTR averages. The most abundant changes of CR intensity are of low amplitude, not exceeding couple percent, and are buried in the CR noise itself. Possible connection of these fluctuations with fluctuations in cloudiness could eventually be revealed only by cross-correlation analyses with the universe of DTR patterns, what is a demanding task, not yet performed. We are thus left with relatively small number of well defined CR intensity changes (both the decreases (FD) and the increases (GLE)), which almost exponentially decrease in number with increasing amplitude, which would hopefully cross the noise threshold and give the signal of the connection, if it is really there. Now, for the days around the day when non-statistical change of CR intensity sets in ("day 0") we average for each day the deviations of actual DTR values from their expected values for all the 210 stations (for some of them the data are missing for some dates of interest). For FD analyses we restrict to only 10 so-called quiet CR days before the effect takes place, since longer periods often contain some other variation of intensity, and, for the same reason, to only 10 days after the beginning of the effect, Finally, we overlap these deviations for all the events chosen and average them on the day-by-day basis. This is what we present in graphic form as final results of our analyses. Fig.1, is a modified version of the figure from our earlier work, where we add, for visual clarity, the true shapes of the most intense FD events along with the corresponding departures of DTR values from their expected values. The presented error bars are external standard errors of the mean at every day's point, and their estimates can be improved primarily by increasing the number of independent deviations from the expected DTR values. Alternatively, these errors may be estimated from the real scatter of noise data points outside the region where the effect of CR intensity changes is assumed to take place (e.g. days –10 to 0 and days 8, 9, 10, in Fig.1). Minding that these are crude estimates, the two agree satisfactorily. We find that for the decreases of CR intensity (FD) higher than about 7% there exists a correctly time correlated and moderately significant increase of DTR, while for smaller changes of CR intensity the deviations from expected DTR by all criteria sink into the noise. For a small sample of FDs higher than 10% (19 events altogether) the effect is most pronounced, and from our Fig.1 one can deduce that it is significant at about the $4\sigma$ level. As seen, the size of the effect shows some proportionality to the intensity of the Forbush decrease, though the significance of this inference is quite low.

We now briefly comment on the criticisms put forward in recent papers by Laken et.al.[15]. and by Erlykin and Wolfendale[16]. These authors managed to rather closely reproduce our results for the FDs, but also demonstrated that in a much wider interval around the "day 0" similar excursions from the expected averages occur frequently enough, seemingly uncorrelated with the well defined CR intensity changes. This is a useful knowledge, which certainly diminishes the significance that we deduced above on the basis of the narrower sample of noice. The objection, however, remains that by taking a much wider interval, which certainly encompasses the non-quiet CR days, the possibility of other deterministic effects to creep in increases, thus possibly increasing the thereafter estimated width of noise beyond its realistic value. Also, we object to the procedure of smoothing used in both the works of Laken et. al. and Wolfendale and Erlykin – smoothing may be useful for purposes of presentation, but it imparts the original statistical properties of non-stationary data in a hard to quantify way, and consequently does not allow for further clear statistical interpretation. Again, while analyzing the periodicities in the DTR series in the search for those that might match the periodicities in the CR data series Laken et. al. resort to the subtraction of the two series, what again changes the statistical properties of the original time series completely (for instance, the result of subtraction of two Poissonian series cannot be Poissonian since negative values appear). We thus rather keep the narrow interval around the CR effect containing only the quiet CR days, and pay the price by the increase of statistical errors due to the small number of points included into the definition of noise properties and possibly by the lowering of their estimates due to the same cause. Nevertheless, the information that excursions of comparable amplitudes of DTR values from the expected averages occur also with no apparent connection with cosmic-ray variations is a valuable one and must be bared in mind while estimating the significance of our results. As will be seen in what follows, the cases of CR intensity increases offer independent evidence in favor of the correlation, of comparable significance. And, independent of the method of analysis, there is apparent proportionality of the amplitude of the effect and the intensity of the CR change in question.

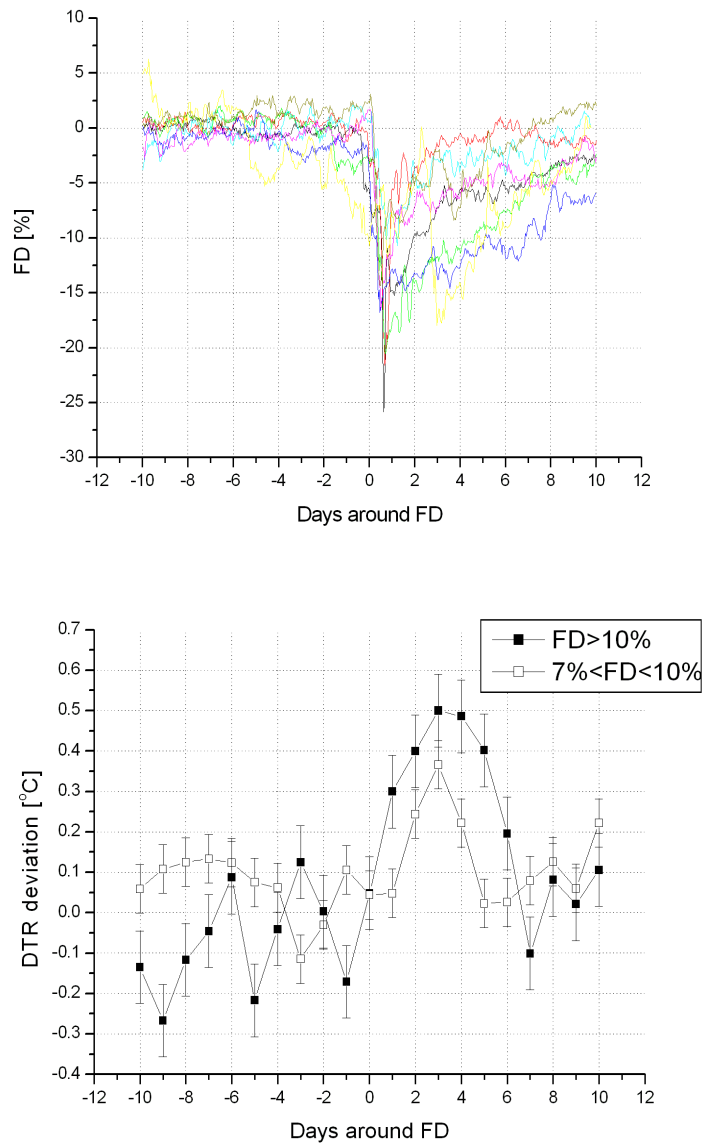

**Fig.1** The results of our previous analysis of the departures from the expected DTR values for the period of 20 days, centered around the day when the Forbush decreases of different amplitudes set in (lower picture). Lower amplitude corresponds to FD events with amplitudes between 7 and 10%, while the higer one correspondes to FD events with amplitudes greataer than 10%. Based on the estimates of intrinsic statistical errors marked on individual points, as well as on the width of the underlying noice estimated from the quiet CR days that presede the effect, this later peak is significant at the 4σ level. Peaks of similar amplitude, however, according to [15] and [16], occur haphasardly in intervals much wider around the "day 0". The effect, if real, reaches its maximum 2 or 3 days after the change of CR intensity begins, and is proportional to the intensity of the change. The delay seems reasonable, minding that it must take considerable time for the effects of decreased ionization to manifest on cloudiness, and then on the daily temperature range. Typical changes of cosmic-ray intensity that correspond to these DTR deviations are, for purposes of illustration, presented in the upper part of the figure.

## 2. The cases of the increased CR intensity and the latitude effect

The sudden non-statistical increases of CR intensity, known as the Ground Level Enhancements (GLE), are of very different time structure than the Forbush decreases. While the FD-s last typically days, with a characteristic fast decrease and a slow recovery of the CR intensity, the GLE-s last typically hours, but with amplitudes generally much higher than the FD-s. The total power in the signal may, however, for those of highest intensity, equal the total power in FD events. Of interest to the present analysis is the fact that they usually do not occur during the quiet CR days, and that they are rarely found well isolated from other non-statistical variations of intensity. This is why, in order not only to keep the information as visually clean as possible, we had to narrow the intervals around the GLE-s to only three days before and after the effect sets in, as compared to 10 days for FD-s. The results for GLE-s of different intensity, as seen by the OULU CR observatory for the period from 1964 to 2010, are presented in Fig.2.

As expected for the genuine effect, it is seen that there is a correctly time correlated decrease of DTR values from their expected values, the magnitude of the DTR decrease being roughly proportional to the magnitude of the CR intensity increase. Also, as might be expected for the increase of ionization in the atmosphere, the changes of DTR following CR intensity increases virtually experience no delay, and, what might be interesting, even seem to manifest sort of precursor effect. Due to the impossibility to define the stationary level of noise from such a narrow time interval with any significance, the statistical analysis we performed for FD-s was not possible to repeat here. It is, however, plausible to assume that the noise in the expected DTR values is on the average constant, and that it is characterized by the same parameters everywhere. Under this assumption one sees that the significance of the data for the most pronounced cases of GLE greater than 30%, is about the same as in the case of FD-s greater than 10%.

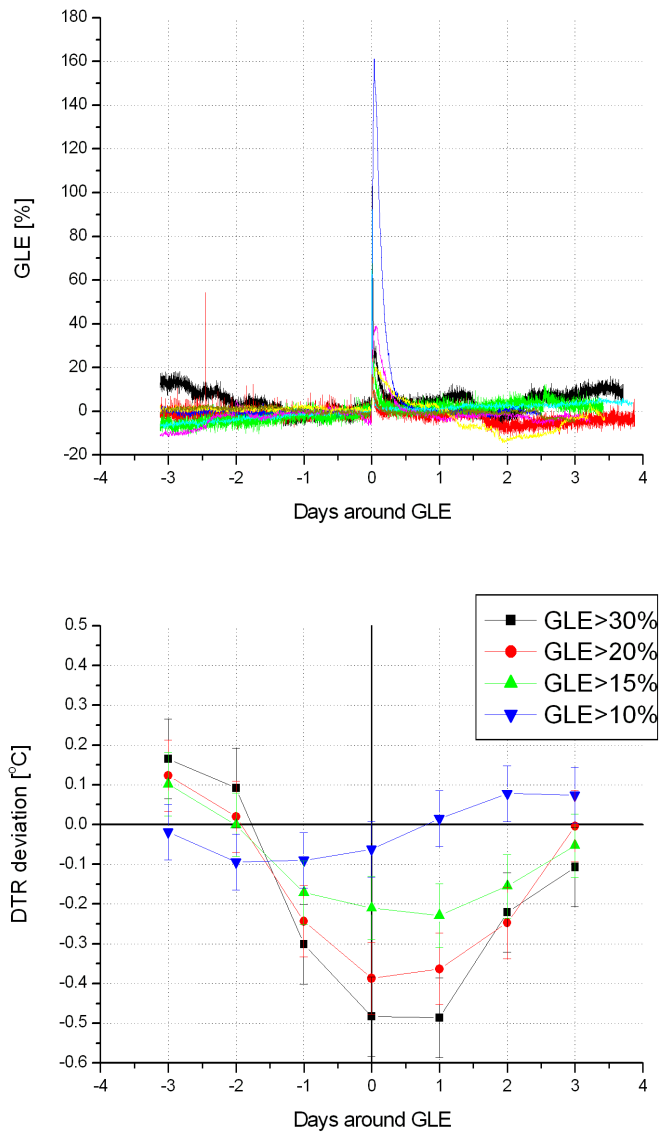

**Fig. 2.** Average deviations of DTR values from their expected average values for the days around the appearance of the GLE effects, for different GLE amplitudes (lower figure). Note the difference in time scale as compared to that in Fig.1. Existence of threshold and proportionality of amplitudes is present as in the case of FDs. Phase relation again seems plausible – no appreciable delay between the increase of the condensation inducing agent and the increase of condensation is expected. Typical high-amplitude increases of cosmic-ray intensity that correspond to these deviations of the DTR values from their expected averages are, for purposes of illustration, presented in the upper part of the figure.

Finally, the DTR deviations that correspond to Forbush decreases bigger than 10%, which are the only ones to allow for this kind of analysis, apparently exhibit some latitude effect (Fig. 3). Minding the sizes of internal statistical errors, which are quoted at the 68% CL, the effect is again of moderate statistical significance.

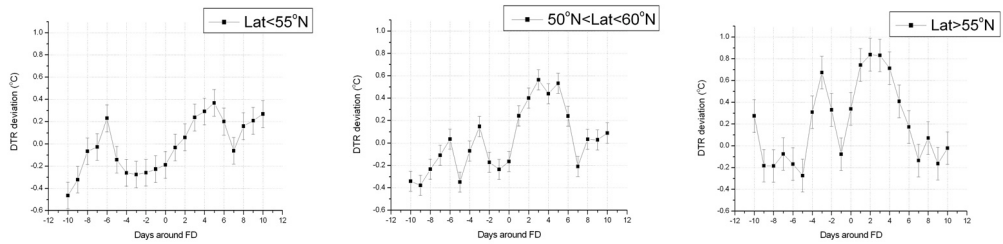

**Fig.3**. Latitude effect in DTR deviations for Forbush decreases bigger than 10%, as seen by the European meteorological stations, a) 111 stations at latitudes below 55°N, b) 106 stations at mid-latitudes between 50°N and 60°N, c) 92 stations at latitudes above 55°N

This is perhaps the place to note that the most numerous well-isolated cosmic-ray effects are those of small amplitudes, which apparently do not produce effects on the DTR that significantly protrude out of the noice, while the number of those with higher amplitudes that seem to exhibit observable effects decreases practically exponentially. This distribution is inherent to the phenomenon and, as is the case with studies of many natural phenomena, one must live with it – statistics is necessarily poor where the effect, if any, is expected to be the greatest. It is only with passage of time and the consequential enrichment of our databases that our inferences will gain more statistical weight. However, our findings even at this stage seem to suggest some connection between cosmic rays and climate. It appears that the transient phenomena on the Sun induce observable climate changes on the Earth both via the minor changes of solar irradiance (*non-ionizing radiations*) and the corresponding and much bigger changes of CR intensity (*ionizing radiations*) [4,10,11], in proportion that still has to be determined.

**Conclusions**

Finally, we sum up the information contained in graphical form in our figures. Four qualitative inferences follow from the present analysis. **First**, the deviations of the DTR from the expected values are positive for the decreases of the CR flux (the Forbush decreases), while, again as expected for the genuine correlation, they are negative for the increases of the CR flux (the GLE effects). **Second**, there is proportionality between the intensity of the CR flux change and the DTR deviation from the expected value. There is also a threshold value of the CR intensity change, below which the correlation is buried in the noise. With present statistics neither the character of the proportionality nor the height of this threshold value are possible to determine quantitatively with sufficient significance. **Third**, phase relations between the CR changes and the DTR responses seem plausible; it should indeed take longer for the state of the atmosphere to react to the decrease of the ionization-inducing agent, than to its increase. **Fourth**, there is a latitude effect of the expected sign – the deviations of the DTR values get more pronounced at higher latitudes. At this level of significance nothing can be said about the causality of the effects on the basis of any of the single correlation, but taken together, they make the inferences much more significant, though difficult to quantify to what extent.

Present analysis cannot be considered conclusive, but is only another piece of evidence that exist in the present-day databases. All four inferences, at their individual modest levels of significance, consistently support the possibility that there is a genuine correlation between the cosmic-ray intensity and the diurnal temperature range, or indirectly, the cloudiness. This fourfold coincidence of independent pieces of evidence has small chances to be accidental. It suggests the common cause to all these correlations. Our conclusions agree in part, and perhaps not only qualitatively, with those of B.H.Brown[14], obtained from a rather different type of analysis. More quantitative conclusions would have to wait for the improvement of statistics. This would, however, come only with the passage of time, when the number of known changes of cosmic-ray intensity of sufficient intensity to overcome the noise is significantly increased. As the final remark we recall that the phenomena on the Sun mostly influence the comparatively low energy portion of the CR spectrum, while the higher energy part, which is responsible for the ionization of the deep athmosphere, remains largely unchanged, what might account for the weakness of the correlation, even if there is a causal relation between the two.

This work is supported by the Ministry for Education, Science and Technological Development of Serbia under the Project OI 171002.